\documentclass{PoS}

\usepackage{amsmath}
\usepackage[textsize=scriptsize]{todonotes}
\let\marginpar\oldmarginpar 
\usepackage{setspace}
\usepackage[mode=buildnew]{standalone}
\usepackage{wrapfig}

\newlength\Colsep

\newcommand{\dif}{\text{d}}

\newcommand{\fdel}[2]{\frac {\partial #1}{\partial #2}}

\newcommand{\intIn}[2]{\int_{#2}\!\dif #1\,}
\newcommand{\intInFT}[3]{\int_{#2}^{#3}\!\dif #1\,}

\newcommand{\tr}{\mathrm{tr}}
\newcommand{\diag}[1]{\mathrm{diag}\!\left(#1\right)}
\newcommand{\trlong}[1]{\tr{\left[{#1}\right]}}

\DeclareMathOperator{\ex}{\mathrm{e}}

\newcommand{\e}[1]{\ex^{#1}}

\newcommand{\order}{\mathcal{O}}

\newcommand{\obs}[1]{\langle #1 \rangle}

\title{Applying recursive numerical integration techniques for solving high dimensional integrals}

\ShortTitle{Recursive numerical integration techniques}

\author{Andreas~Ammon$^e$, Alan~Genz$^b$, Tobias~Hartung$^c$,
  Karl~Jansen$^a$,
  Hernan~Le\"ovey$^d$, \speaker{Julia~Volmer}$^a$\\ \\
\llap{$^a$}NIC, DESY\\ Platanenallee 6, D-15738 Zeuthen,
Germany\\ \\
\llap{$^b$}Department of Mathematics, Washington State University\\
  Pullman, WA 99164-3113 USA\\ \\
\llap{$^c$}Department of Mathematics, King's College London\\ Strand,
London WC2R 2LS, United Kingdom\\ \\
\llap{$^d$}Institut f\"ur Mathematik, Humboldt-Universit\"at zu
Berlin\\ Unter den Linden 6, D-10099 Berlin\\ \\
\llap{$^d$}IVU Traffic Technologies AG\\ Bundesallee 88, 12161 Berlin, Germany\\ \\
Email: \email{andreas.ammon@desy.de}, \email{genz@math.wsu.edu},
\email{tobias.hartung@kcl.ac.uk}, \email{karl.jansen@desy.de},
\email{leovey@math.hu-berlin.de}, \email{julia.volmer@desy.de}
}

\abstract{
The error scaling for Markov-Chain Monte Carlo techniques (MCMC) with
$N$ samples behaves like $1/\sqrt{N}$. This scaling makes it often
very time intensive to reduce the error of computed observables, in
particular for applications in lattice QCD. It is therefore highly
desirable to have alternative methods at hand which show an improved
error scaling. One candidate for such an alternative integration
technique is the method of recursive numerical integration (RNI). The
basic idea of this method is to use an efficient low-dimensional
quadrature rule (usually of Gaussian type) and apply it iteratively to
integrate over high-dimensional observables and Boltzmann weights. We
present the application of such an algorithm to the topological rotor
and the anharmonic oscillator and compare the error scaling to MCMC
results. In particular, we demonstrate that the RNI technique shows an
error scaling in the number of integration points $m$ that is at least exponential.
}

\FullConference{34th annual International Symposium on Lattice Field Theory\\
         24-30 July 2016\\
         University of Southampton, UK}

\begin{document}

\section{Introduction}
For evaluation of the high dimensional path integrals in numerical
simulations of statistical physics and lattice-QCD \cite{Luscher:2010ae} mainly
Markov-Chain Monte Carlo methods (MCMC) are used. 

If we have a model with discrete variables
$\{x_1, ..., x_d\}, x_i \in D \subset \mathbb{R}$ and an action
$S[x] \equiv S(x_1, ..., x_d)$, we can measure the expectation value
of an observable $O[x]$ in this model by
\begin{align}
  \obs{O} = \frac
  {\intIn{x}{D^d} O[x] \, \e{-S[x]}}
  {\intIn{x}{D^d} \e{-S[x]}}.
  \label{equ:obs}
\end{align}
These integrals are usually impossible to approximate with
reasonable sample sizes by means of direct sampling because $d$ is large. MCMC methods can
be applied to large dimensions $d$ because they often choose $N$
sampling configurations $[x]$ from a probability distribution
dependent on $\e{-S[x]}$. On the other side the
error of $\obs{O}$
scales weakly, namely like$\frac{1}{\sqrt{N}}$. 
Additionally, the consecutively chosen sampling points have a
dependence and these autocorrelations can lead to
large errors since they necessitate very long MCMC runs.
Therefore it is highly desirable to look for
alternative methods to improve the weak error scaling, especially for very
time-intensive computations as they are done for example in
lattice-QCD.

One alternative approach, the Recursive Numerical Integration (RNI)
method \cite{Genz86, Hayter06}, promises to overcome both MCMC
drawbacks, at least in principle. RNI is a polynomially exact method,
not of statistical nature,
therefore we directly avoid autocorrelations and the theoretical
predictions for the error scaling are usually at least exponential. In this
method we compute numerator and denominator in \eqref{equ:obs}
separately. Similar to MCMC we also exploit the weighting function
$\e{-S[x]}$ but not as a probability density but by using its
structure to simplify the integrals and finally apply Gaussian
quadrature to solve them numerically to get a polynominally exact
solution. The simplification of course depends a lot on the action
itself and therefore is model dependent. Here we apply the method to
the topological oscillator to obtain a first test whether the
predicted improved error scaling can indeed be reached.

\section{Recursive Numerical Integration}
\begin{figure}
  \includegraphics[width=1.\textwidth]{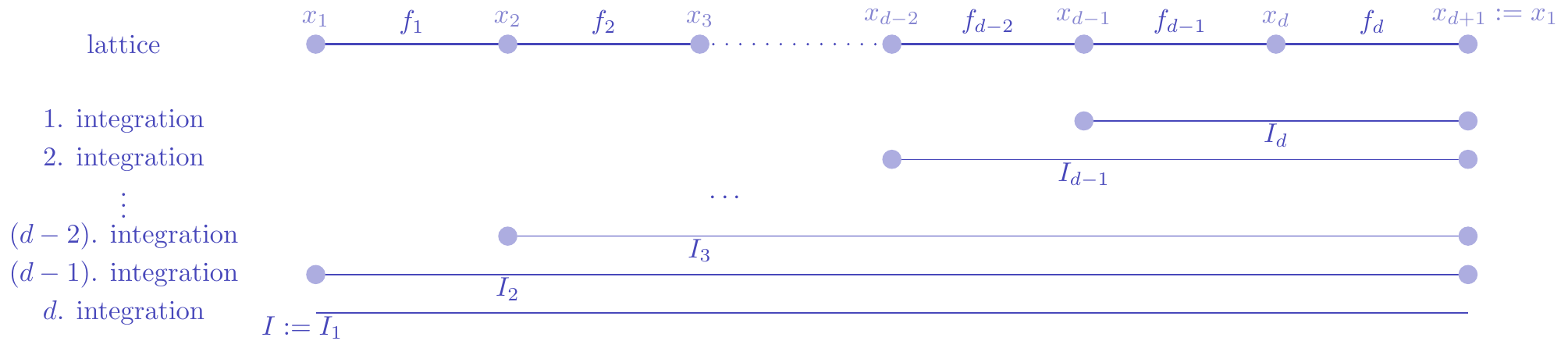}
\caption{Visualization of the recursive integration steps for a
  1-dimensional lattice with $d$ points $x_i$, next-neighbor couplings
  $f_i$ and periodic boundaries: How to consecutively calculate $I_i$,
  $i \in \{d, d-1, ..., 1\}$ by integrating out the lattice variable
  $x_i$ (shown in the first row) and finally arrive at the desired
  integral $I$.}
\label{fig:integrandStructure}
\end{figure}

We use the principle of Recursive Numerical Integration (RNI)
\cite{Genz86, Hayter06} to derive the final formula we use to compute each
of the two main integrals in \eqref{equ:obs}. The derivation consists of three
main steps. First we identify a structure in the integrand,
secondly we integrate recursively, and finally choose a quadrature
rule to perform each integration numerically. The last paragraph in this section
is dedicated to the discussion of the differences in computing the two
integrals in \eqref{equ:obs}
\paragraph{Integrand Structure}
Throughout the whole paper we look at physical 1-dimensional models,
schematically shown at the top of figure \ref{fig:integrandStructure},
with $d$ lattice points, only next-neighbor couplings $f_i$ (which are
in general different from each other) and
periodic boundaries. Because in addition all considered
observables are of algebraic nature we can write each of the two
integrals in \eqref{equ:obs} in the form
\begin{align}
  I = \intIn{x_1}{D} ... \intIn{x_d}{D}  \prod_{i=1}^d f_i(x_i,
  x_{i+1}).
  \label{equ:intNotOrdered}
\end{align}

\paragraph{Recursive Integration}
We can reorder expression \eqref{equ:intNotOrdered}. Because of the
next-neighbor coupling every lattice point $x_i$ appears only twice,
in $f_i$ and $f_{i-1}$, and $I$ can be written
\begin{align}
  \resizebox{.9\hsize}{!}{$\displaystyle{
  I = \underbrace{\intIn{x_1}{D} ... \underbrace{\bigg(\intIn{x_{d-2}}{D} f_{d-3}(x_{d-3},
  x_{d-2}) \cdot \underbrace{\bigg(\intIn{x_{d-1}}{D} f_{d-2}(x_{d-2},
  x_{d-1}) \cdot \underbrace{\bigg(\intIn{x_d}{D} f_{d-1}(x_{d-1}, x_d)
  \cdot f_d(x_d, x_{d+1}) \bigg)}_{I_d} \bigg)}_{I_{d-1}}
  \bigg)}_{I_{d-2}}}_{I_1}
  }$}.
  \label{equ:intOrdered}
\end{align}
This recursive integration process is visualized in the lower part of figure
\ref{fig:integrandStructure}. The first integration we can perform is the
integral standing at the right end of equation \eqref{equ:intOrdered}
over $f_{d-1}$ and $f_d$. Here we integrate out $x_d$ and 
call the resulting integral $I_d$. $I_d$ has two open indices, $x_{d-1}$
and $x_{d+1}$. It is visualized in the figure as an
uninterrupted line including $f_{d-1}$ and $f_d$ from knot $x_{d-1}$ to
$x_{d+1}$.

In the second integration step we integrate out $x_{d-1}$. Here we
integrate over the already calculated $I_d$ and the new $f_{d-2}$,
which, as $I_d$, depends on $x_{d-1}$.
This procedure is repeated until we have integrated out all $x_i$,
where for the last integration over $x_1$ we just have to integrate
over $I_2(x_1, x_1)$. 

\paragraph{Numerical Integration}
For numerical integration we want to approximate each of the above
integrals with some integrand $g(x)$ by a weighted sum
(quadrature)  $\intIn{x}{D} g(x) \approx \sum_{r=1}^m w^r g(x^r)$,
where we evaluate the integrand at specific mesh points $x^r$ and
weight this result by corresponding weights $w^r$.\footnote{The superscript
here is used to distinguish these integration indices from
the discretization indices and should not be
interpreted as some exponential.}
Applying this approximation to each integral in \eqref{equ:intOrdered}
leads to a transformation of
$\intIn{x_i}{D} \rightsquigarrow \sum_{r=1}^m w_i^r$ and
$f_i(x_i, x_{i+1}) \rightsquigarrow M_i(x_i^j, x_{i+1}^k) =: M_i^{j,k}$.
We apply this approximation to the first integration step to get
\begin{align}
  I_d(x_{d-1}^s, x_{1}^t) &\approx \sum_{r=1}^m w^r \, M^{s,r}_{d-1} \,
                              M^{r,t}_d
                            = \bigg( \underbrace{M_{d-1} \cdot
                              \diag{w^1,w^2,...,w^m}}_{\widetilde{M}_{d-1}} \cdot M_d \bigg)^{s,t},
\end{align}
which is just the $(s,t)$ entry of the matrixproduct in brackets. This is
valid for all integration mesh points of $x_{d-1}^s$ and $x_{1}^t$,
$s, t \in \{1, ..., m\}$.

For each of the following integration steps $i \in \{2, ..., d-1\}$, we get an additional sum, weight
and a matrix, which gives an additional $\widetilde{M}_{d-i+1}$ in the
matrix product. Therefore after repeating this step $d-2$ times we arrive at
  $I_2(x_1^t, x_1^t) \approx \left( \left( \prod_{i=1}^{d-1}
   \widetilde{M}_i \right) M_d \right)^{t,t}$.
Integrating finally over $x_1$ gives
\begin{align}
  I = I_1 = \intIn{x_1}{D} I_2(x_1, x_1) \approx \sum_{t=1}^m w^t
  I_2^{t,t}
  = \trlong{\diag{w^1, ..., w^m} I_2^{t,t}}
  = \trlong{\left( \prod_{i=1}^{d}
   \widetilde{M}_i \right)}.
  \label{equ:finalInt}
\end{align}

Because RNI is a deterministic method and we are mostly interested in
the scaling of the error and not the value of the integration itself, we estimate the
error of the integral $I(m)$ at some number of mesh points $m$ by a
truncation error, i.e. by computing $I$ at a larger $m_0>m$ value and
compute the difference
  $\Delta I(m) = |I(m) - I(m_0)|$.

At this point, we have not yet specified which quadrature to use. For every 1-dimensional integration
$2m$ parameters, the mesh points and weights, have to be determined. By using
Gauss quadrature we approximate the integrand by a polynomial of
degree $2m-1$ \footnote{which has $2m$ free coefficients} and therefore obtain a
polynomially exact solution to our integral.
One valid choice for these
polynomials are the Legendre polynoms of degree
$m$. This is a good choice because the functions we look at are
mostly exponentials which are approximated well by Legendre polynomials.

If the integrand $g(x)$ is not itself a polynomial of
degree $2m-1$ but sufficiently smooth in the integration range $D$,
$g(x) \in C^{2m}(D)$, the error scaling of the integral approximation
by using Gauss quadrature is asymptotically (for large enough $m$)
\begin{align}
  \order{\left(\frac{1}{(2m)!}\right)} \sim \order{\left(
  \frac{\e{2m}}{\sqrt{2\pi 2m} (2m)^{2m}} \right)} = \order{\bigg(
  \exp{[-2m\ln{m}]} \cdot
  \frac{1}{\sqrt{m}} \bigg)},
  \label{equ:errorGauss}
\end{align}
where we used the Stirling formula to estimate the factorial.

\paragraph{Types of integrals}
There are two types of integrals we want to compute, the numerator
and the denominator of equation \eqref{equ:obs}. The denominator
integrates the Boltzmann weight. Assuming isotropy of a given model we
obtain here $f_1 = f_2 = ... = f_d$.
Therefore also $M_1 = M_2 = ... = M_d$ and the final integral
\eqref{equ:finalInt} simplifies to $I \approx
\tr [(\widetilde{M})^d]$. This can be evaluated by either computing the
eigenvalues of $\widetilde{M}$, raise them to the $d$th power and sum
them up or computing the $d$th power of the matrix explicitly and
take the trace. The choice of the
approximating polynomial and the chosen quadrature rule are very important since
$\widetilde{M}$ is a $m \times m$ matrix. Hence, the smaller $m$ is,
the easier it is to compute $I$.

The numerator depends strongly on the observable. This is normally
some summation and/or multiplication of the variables $x_1,..., x_d$. 
In general this means we obtain different $f_i$'s though most of them
will still coincide in practice. Thus, we need to compute the trace
of a product of at most $d$ matrices, some of which may be raised to a
power less than $d$.

\section{Topological Oscillator}
In this section, we apply the discussed method of RNI to the topological oscillator.
This simple physical 1-dimensional model shows some characteristic features of non-linear
$\sigma$-models and gauge theories, see e.g. \cite{Rothe:1992nt,Montvay:1994cy,Gattringer:2010zz}.
It describes a particle with mass $M$ moving on a circle with radius
$R$. The basic degree of freedom is an angle $\phi$ dependent on time $t$. The
action of this system is the integration of the kinetic energy of the
particle over a time period $T$.
We discretize one time period in $d$ slices with spacing $a$. On each
timeslice lives an angle $\phi_i = \phi(t_i), i \in \{1, ..., d\}$.
\begin{align}
  S_{\text{continuum}}(\phi) = \frac{MR^2}{2} \intInFT{t}{0}{T}
  \left(\fdel{\phi}{t}\right)^2 
  \quad \stackrel{\text{lattice}}{\Longrightarrow} \quad
  S[\phi] = \frac{MR^2}{a} \sum_{i=1}^d (1 - \cos (\phi_{i+1}-\phi_i) ).
\end{align}
The right side of the above equation is one possibility for the discretized action.
One characteristic quantity of this system is the topological charge.
This is the number of complete revolutions of the rotor in the time
period $T$,
\begin{align}
  Q_{\text{continuum}}(\phi) = \frac{1}{2\pi} \intInFT{t}{0}{T} \left(
  \fdel{\phi}{t} \right)
  \quad \stackrel{\text{lattice}}{\Longrightarrow} \quad 
  Q[\phi] = 
  \frac{1}{2\pi} \sum_{i=1}^d (\phi_{i+1} - \phi_i) \bmod[-\pi, \pi).
\end{align}
A more physical observable to characterize a system is the width of
the distribution $Q$ normalized by the time period $T$, the
topological susceptibility $\chi$, with the continuum limit
\begin{align}
  \chi[\phi] = \frac{\obs{Q^2[\phi]}}{T} \xrightarrow[a \cdot d =\text{const}]{a \rightarrow 0, d
  \rightarrow \infty} \frac{1}{4\pi^2MR^2}.
  \label{equ:chi}
\end{align}

\section{Numerical Results}
As a benchmark quantity we compute the topological susceptibility
with the RNI method. We first test the correctness of the method by
checking that we obtain the right correct continuum limit value. Then
we look at the error scaling as a function of the number of
integration steps we use. And finally we compare the error scaling and
the cost of the computation with the optimal MCMC Cluster algorithm
results by looking at the runtime of both methods on a standalone
computer.

\begin{figure}[h]
\begin{minipage}[c]{0.48\textwidth}
  \centering
  \includegraphics[width=.9\textwidth, page=4]
  {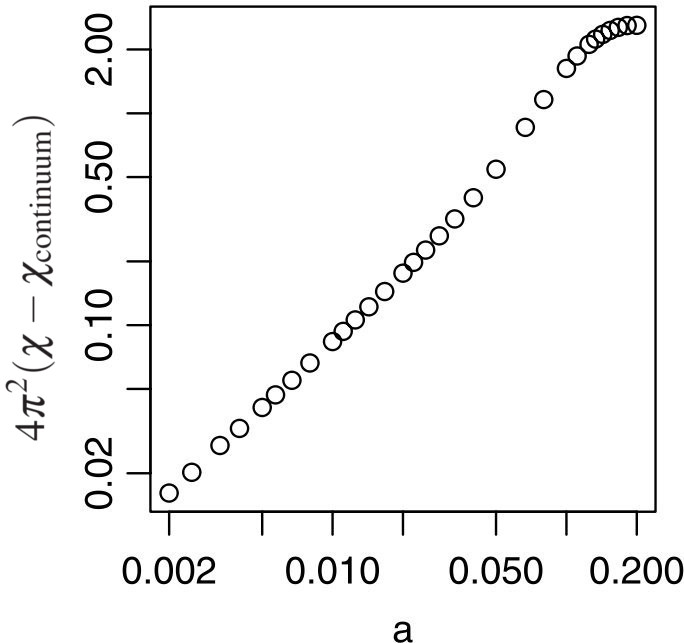}
  \caption{Continuum extrapolation of the topological susceptibility
    $\chi$ of the topological oscillator, computed with RNI.}
\label{fig:gaussobservables}
\end{minipage}\hfill
\begin{minipage}[c]{0.48\textwidth}
  \centering
  \includegraphics[width=.9\textwidth]
  {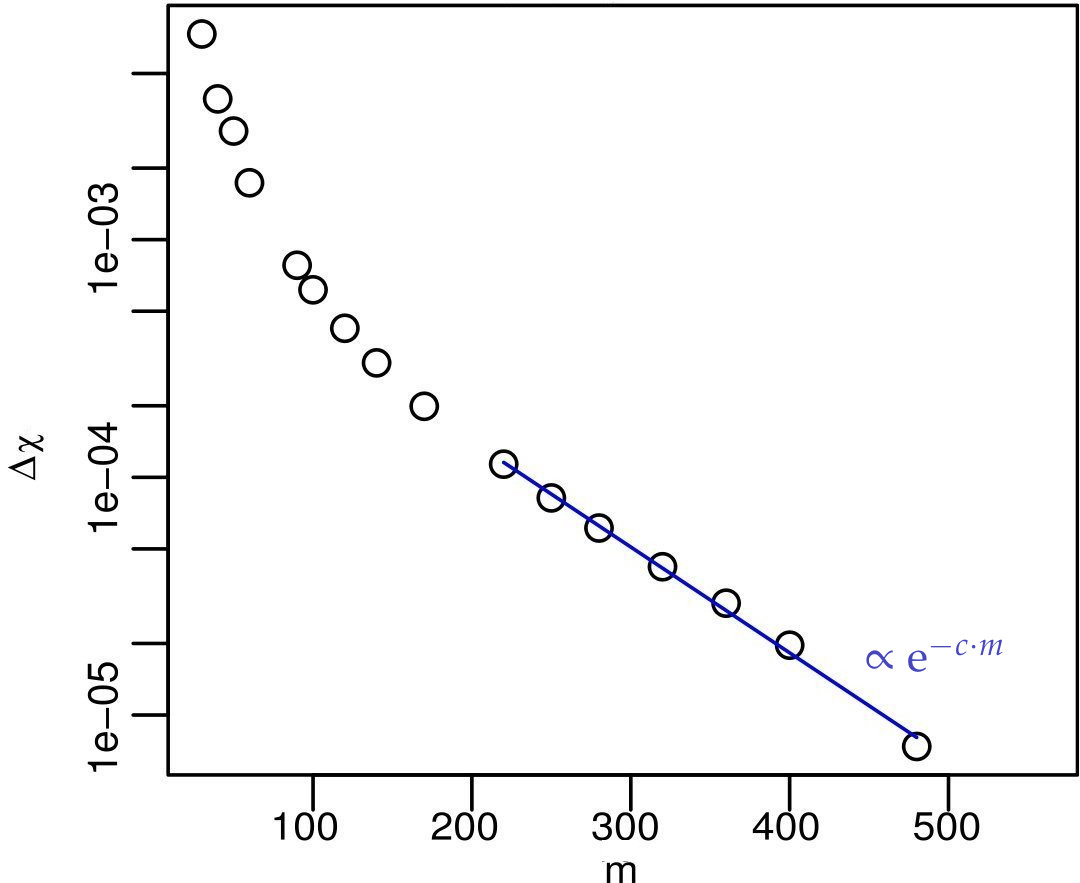} 
  \caption{Error scaling of the topological susceptibility
    $\Delta \chi$ computed with RNI with number of integration points
    $m$. The blue line is an exponential fit in the range
    $220 \le m \le 480$.}
  \label{fig:gauss-dchi-m}
\end{minipage}
\end{figure}

To demonstrate the correctness of the results computed with RNI we
first fix the number of integration points to $m = 120$ and $MR^2 = 0.25$ and computete the topological susceptability
$\chi$ for different lattice spacings $a$. The expected continuum
limit value, considering \eqref{equ:chi}, is $\chi_{\text{continuum}} = \frac{1}{\pi^2}$. The result
of the computed $\chi$
subtracted by this expected theoretical value is shown in figure
\ref{fig:gaussobservables} and, as expected, it converges to zero at
$a=0$, compare also \cite{Bietenholz:2010xg}.

Next we examine the error scaling of the RNI method for which we
use the truncation
error as discussed above. We choose $MR^2=0.25$, $a=0.4$ and a $\chi$
gauge value at $m_0 = 560$ to compute the truncation error
for $\chi$ values at $m < 500$. Results are shown in
figure \ref{fig:gauss-dchi-m}.
We fit an exponential $\exp{(-cm)}$ to the data points in the range
$220 \le m \le 480$, which appears as a straight line in figure
\ref{fig:gauss-dchi-m}, where a logarithmic scale is used. The good
agreement between the data and the exponential suggests that
asymptotically the error scales at least exponentially fast. This is
comparable with the expected error scaling for Gauss quadrature in
\eqref{equ:errorGauss}. The expected weakening of the exponential
decay by the additional $\ln{m}$ in the exponent and the $1/\sqrt{m}$
behavior cannot be resolved here, probably because we are not yet
fully in the asymptotic regime where \eqref{equ:errorGauss} holds.

\begin{wrapfigure}[16]{r}{0.5\textwidth}
  \centering
  \includegraphics[width=.5\textwidth]{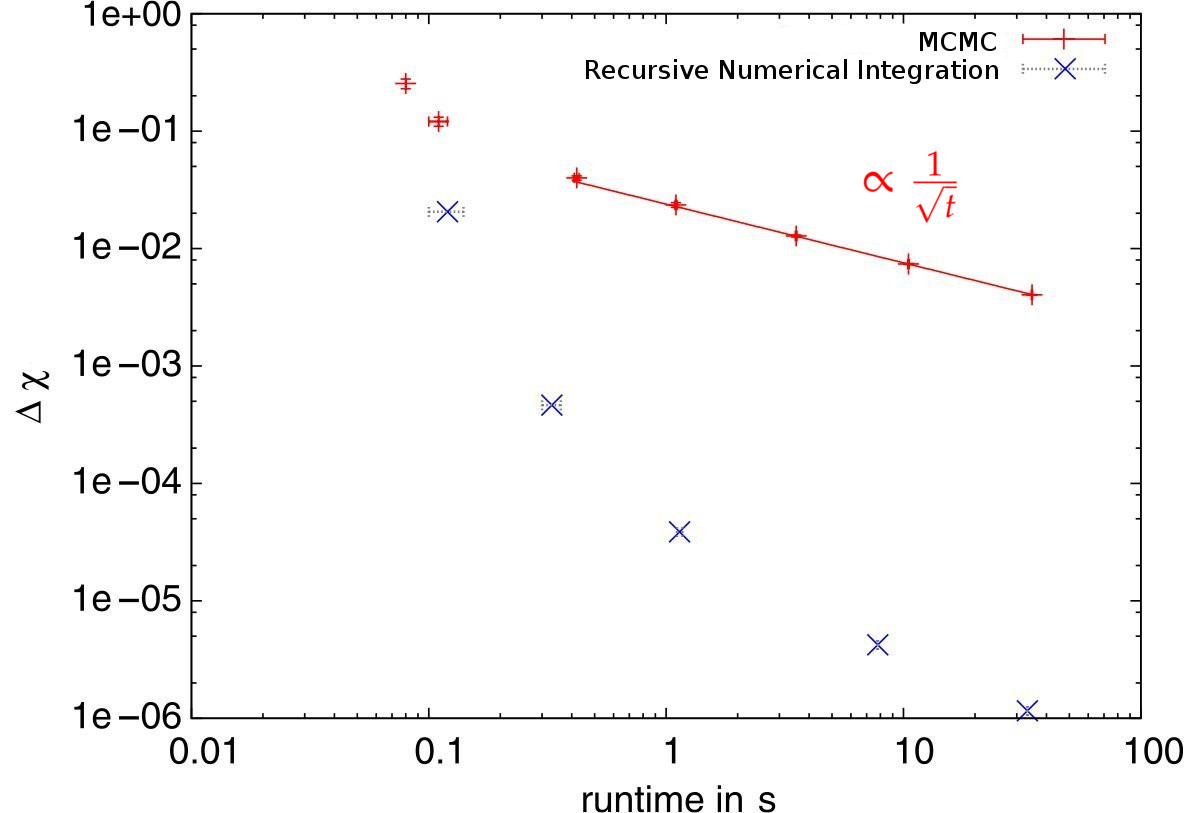} 
  \caption{
    We show the run-time $t$ in
    seconds needed for the Cluster MCMC algorithm and RNI with Gauss
    quadrature to get a given error on the topological
    susceptibility on a 
    stand-alone computer.
    The red line shows the expected $1 / \sqrt{t}$ behavior
    of MCMC.}
  \label{fig:dchi-t}
\end{wrapfigure}

With an exponential error scaling RNI will clearly outperform any
other method with an algebraic error behavior, especially the
$1 / \sqrt{N}$ behavior of MCMC. An interesting question is whether
for smaller, more practical values of $m$ RNI still gives better
results than MCMC. Therefore we compare MCMC and RNI directly. In the
special case of the Topological Oscillator we can apply a specific
kind of MCMC algorithm, the cluster algorithm
\cite{Niedermayer:1996ea}. This algorithm leaves the autocorrelation
time almost constant when going to smaller lattice spacings and hence
is an optimal algorithm for our system. Because both methods use
different error scaling variables (number of mesh points $m$ for RNI
and number of MCMC samples $N$ for the Cluster algorithm) we compare
them by runtime $t$ on a standalone computer. For these computations
we use $a=0.1$ and $MR^2=0.25$ for both methods and vary
$N = 10^2 ... 10^6$ for MCMC and $m = 10 ... 300$ for RNI. The runtime
varies depending on other processes running on the computer, therefore
we repeat every measurement 10 times to get an error estimation of the
runtime. For the cluster algorithm, in addition, the size and
distributions of the generated clusters can vary, leading to different
runtimes of the algorithm. The error on the topological susceptibility
can be estimated for the cluster algorithm by the slightly different
topological susceptibility results of the 10 runs and from its
distribution the error on the error can be roughly determined. For RNI
we get for a fixed $m$ always the same topological susceptibility
result. We estimate the error, as before, by the truncation error,
here with a gauge value at $m_0=400$. The error of this error is
negleced here. Results can be seen figure \ref{fig:dchi-t}. For MCMC
we observe the expected $1 / \sqrt{t}$ behavior, visualized by the red
line in the figure. The RNI error scaling appears to not be
exponential because we are considering values of $m$ that are too
small for the asymptotic error scaling to be observed. However,
although RNI is not yet in that regime, it already outperforms MCMC by
orders of magnitudes.

\section{Conclusion}
In this paper we have applied the method of Recursive Numerical
Integration with Gauss-quadrature to a quantum mechnical system,
namely the topological rotor.
RNI is a method to numerically compute a high dimensional integral.
It uses the structure of the integrand, here the next-neighbor coupling, to
convert the high dimensional integral in many recursively computed small dimensional
integrals which can be solved with high precision by using Gauss
quadrature.

The topological rotor model is simple enough to perform a first test
of the RNI method but it also shows already some characteristic
features of spin systems and even gauge theories. We compute the width
of the distribution of the topological charge, the topological
susceptibility as a benchmark quantity. There we find an exponentially
fast error scaling. Although theoretically the error scaling should be
even faster than exponential, we attribute this finding to the fact
that we still work in an intermediate range of integration points and
are not yet in the asymptotic regime. Comparing RNI directly with the
for this model optimal MCMC Cluster algorithm shows an improvement of
the error for RNI of several orders of magnitude even for a number of
integration points where we are not yet in the intermediate,
exponentially fast error scaling regime. Further applications of RNI
to the anharmonic oscillator are reported in \cite{Ammon201671} and
show extremely good results over a very broad range of parameters.

In symmary, RNI turns out to be an alternative method to MCMC because
it leads to greatly improved error scaling and to order of magnitude
reduced errors for a given runtime. This is crucial, especially
in simulations with a larger number of dimensions.
On the other side, applying RNI to a system with a
larger number of dimensions results in the problem that the
number of next neighbors doubles and instead of matrices $M^{ij}$
we have to deal with tensors $M^{ijkl}$ which makes the whole
computation not feasible any more. Therefore we hope to combine this
method with other techniques which are more suitable for higher
dimensions to exploit the Gauss-quadrature error scaling
to speed up computions.

\bibliographystyle{JHEP_noTitle}
\bibliography{../MainVersion/Article.bib}

\end{document}